\documentclass[prl,twocolumn,showpacs,amsmath]{revtex4-1}

\usepackage{graphicx}

\begin{document}

\title{Cyclotron-resonant exciton transfer between the nearly free and strongly localized radiative states of a two-dimensional hole gas in a high magnetic field}

\author{L. Bryja}\email{Leszek.Bryja@pwr.wroc.pl}
\author{J. Jadczak}
\author{A. W\'ojs}
\affiliation{Institute of Physics, Wroc{\l}aw University of Technology, 50-370 Wroc{\l}aw, Poland}

\author{G. Bartsch}
\author{D.~R. Yakovlev}
\author{M. Bayer}
\affiliation{Experimentelle Physik 2, Technische Universit\"at Dortmund, D-44221 Dortmund, Germany}

\author{P. Plochocka}
\author{M. Potemski}
\affiliation{Laboratoire National des Champs Magn\'etiques Intenses, CNRS-UJF-UPS-INSA, 38042 Grenoble, France}

\author{D. Reuter}
\author{A. D. Wieck}
\affiliation{Lehrstuhl f\"ur Angewandte Festk\"orperphysik, Ruhr-Universit\"at Bochum, D-44780 Bochum, Germany}

\date{\today}

\begin{abstract}
Avoided crossing of the emission lines of a nearly free positive trion and a cyclotron replica of an exciton bound to an interface acceptor has been observed in the magneto-photoluminescence spectra of $p$-doped GaAs quantum wells. Identification of the localized state depended on the precise mapping of the anti-crossing pattern. The underlying coupling is caused by an exciton transfer combined with a resonant cyclotron excitation of an additional hole. The emission spectrum of the resulting magnetically tunable coherent state probes weak localization in the quantum well. 
\end{abstract}

\pacs{71.35.Ji, 71.35.Pq, 73.21.Fg, 78.20.Ls}

\maketitle

Carrier localization is attracting strong current interest fueled by anticipated applications in quantum information technology. For example, charged quantum dots may be used for storage purposes \cite{Kouwenhoven01}, while localized spins are considered as promising quantum bit candidates \cite{Imamoglu99}. Particularly for the latter purpose it seems promising to study also carriers bound to defect atoms \cite{Dutt07}, as smaller extension of their wave function (compared to quantum dots) enhances protection from decoherence. Because of their weaker coupling with the nuclei \cite{Syperek07}, valence band holes appear especially attractive for information storage by localized spins.

On the other hand, small spatial extent also aggravates external manipulation. For example, matrix elements for optical excitation are strongly reduced. In addition, long-range coupling between confined excitations is expected to be rather weak. A possible solution may be coupling of the localized excitations to delocalized ones. This can be obtained by placing the defects close to a quantum well, serving as an interface channel. The quantum well may also be used for efficient optical excitation, after which the photo-injected (and possibly spin-polarized) carriers are exchanged with the localization centers. For deterministic coupling between localized and delocalized excitations they must be brought in resonance where they show a pronounced avoided level crossing -- very similar to the avoided crossing observed for tunnel coupled electronic states in quantum dot molecules \cite{Bayer01}.

\begin{figure}
\includegraphics[width=0.45\textwidth]{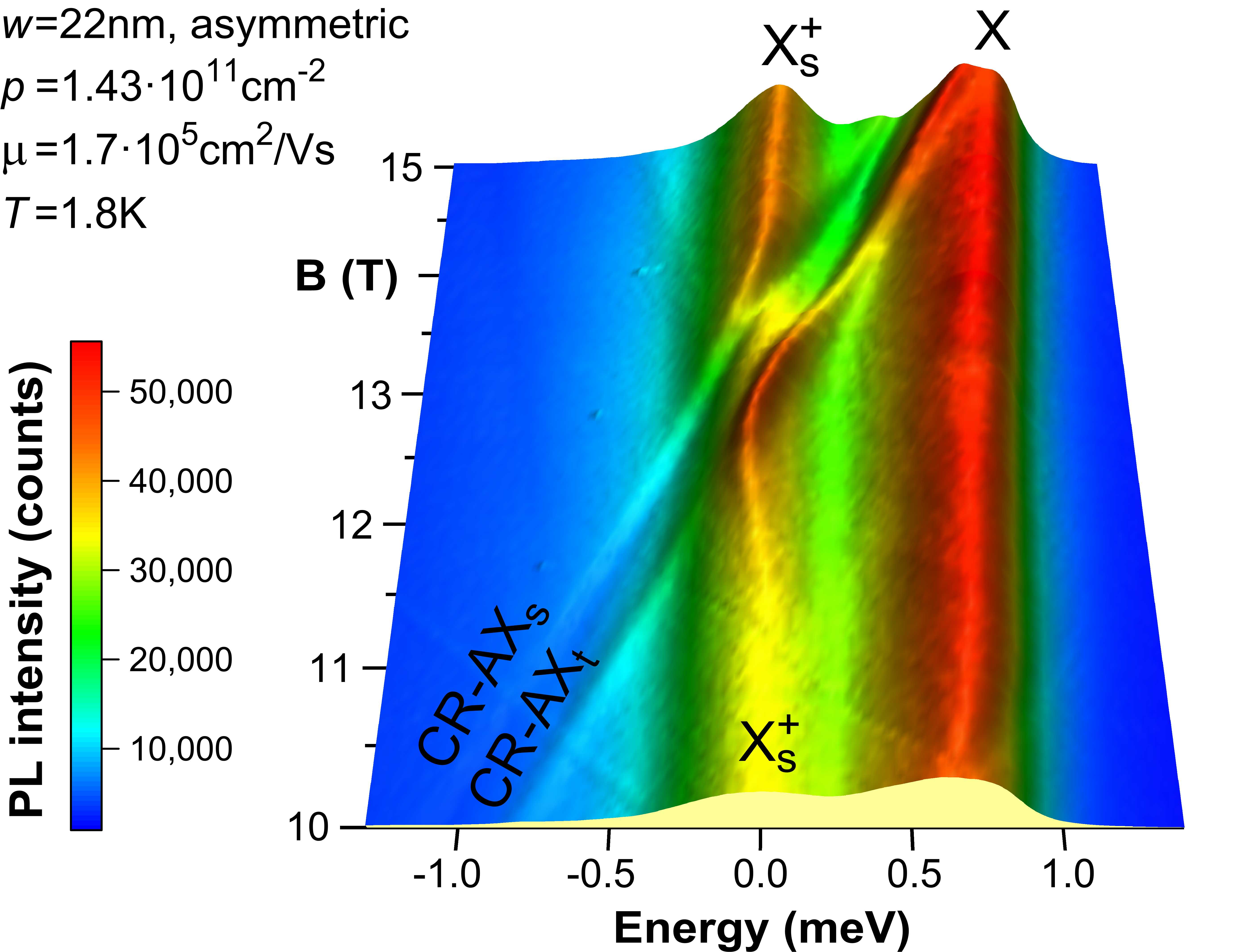}
\caption{(color online) Evolution of the photoluminescence spectrum in the $\sigma^-$ polarization as a function of magnetic field $B$. In fields around $B_c=13.2$~T, avoided crossings between the pair of exchange-split CR-AX lines and the X$^+$ line are clearly observed, demonstrating the coupling via cyclotron-resonant exchange of an electron-hole pair.
  For better visibility, at each magnetic field the energy is linearly shifted to the X$^+$ resonance (slope 0.7~meV/T, constant 1520.8~meV).}
\label{fig1}
\end{figure}

In this Letter, we report on the coupling between the nearly free and strongly localized states of different excitonic complexes formed in a two-dimensional hole gas in a high magnetic field, and study optical emission from the resulting magnetically tunable coherent state. (By ``nearly free'' we mean either mobile or, more likely, weakly localized on remote charges or width fluctuations, in contrast to the ``strong localization'' on nearby charges.) The interacting states are the positive trion X$^+$ \cite{Shields95} and the AX complex whose binding center A$^-$ is a bare acceptor at an interface of the well. In a high magnetic field, the difference in their binding energy is compensated by a cyclotron excitation of the hole gas, enabling their coupling through resonant exchange of an electron-hole pair. This is observed in photoluminescence as an avoided crossing of the emission lines attributed to the X$^+$ and a cyclotron replica (CR) of the AX, as shown in Fig.~\ref{fig1} (discussed further in detail).

In contrast to the previous studies of the shake-up effect \cite{Finkelstein97}, we used a gas of holes instead of electrons. This was motivated by the higher hole mass, placing the replica CR-AX in the suitable energy range for achieving resonance with the X$^+$. On the other hand, in contrast to the previous studies of the hole shake-up of nearly free trions \cite{Glasberg01,Bryja07}, we detect an opposite replica (with cyclotron gap {\em increasing} emission energy) of a {\em localized} complex.

We have examined a selection of superior quality GaAs quantum wells, ranging in width between $w=15$ and 40~nm. They were grown by molecular beam epitaxy on a (001)-oriented semi-insulating GaAs substrate, employing the following growth sequence: 100~nm GaAs, 5~nm AlAs, 200~nm GaAs, 150~nm Ga$_{0.65}$Al$_{0.35}$As, the superlattice consisting of 33 repetitions of 2~nm GaAs and 1~nm AlAs, 57~nm Ga$_{0.65}$Al$_{0.35}$As, $w$-wide GaAs quantum well, 40~nm Ga$_{0.65}$Al$_{0.35}$As, Carbon $\delta$-doping, 80~nm Ga$_{0.65}$Al$_{0.35}$As, and the 5~nm Carbon $\delta$-doped GaAs cap. An exception is the $w=15$~nm well, doped symmetrically on both sides.

In all investigated samples, the hole mobility measured at $T=4.2$~K was nearly the same, $\mu\approx10^5$~cm$^2$/Vs. The hole concentration was accurately estimated from the quantum Hall effect measurements (performed in the van der Pauw configuration, in parallel with photoluminescence). In the dark it varied slightly, in the range $p=(1.2-1.9)\cdot10^{11}$~cm$^{-2}$; under illumination, it decreased linearly with the excitation power density (by less than 10\% in the actual experiments.) 

Photoluminescence was excited by the 632.8~nm line of a Helium-Neon laser, with the photon energy below the band gap of the barrier. The applied laser power density varied from $P=5$ to 20~mW/cm$^2$. The measurements were performed in a bath liquid helium cryostat, at temperatures varying from $T=1.8$ to 4.2~K. Magnetic field was applied in the Faraday configuration; it was changed with a small step $\Delta B=0.05$~T, up to the maximum value $B=22$~T. The fiber optics was used, with a linear polarizer and a quarter-wave plate placed close to the sample. The $\sigma^-$ and $\sigma^+$ helicities were switched by reversing the field direction. The spectra were analyzed using 0.5~m and 1.0~m long monochromators and a liquid Nitrogen cooled CCD camera.

\begin{figure}
\includegraphics[width=0.45\textwidth]{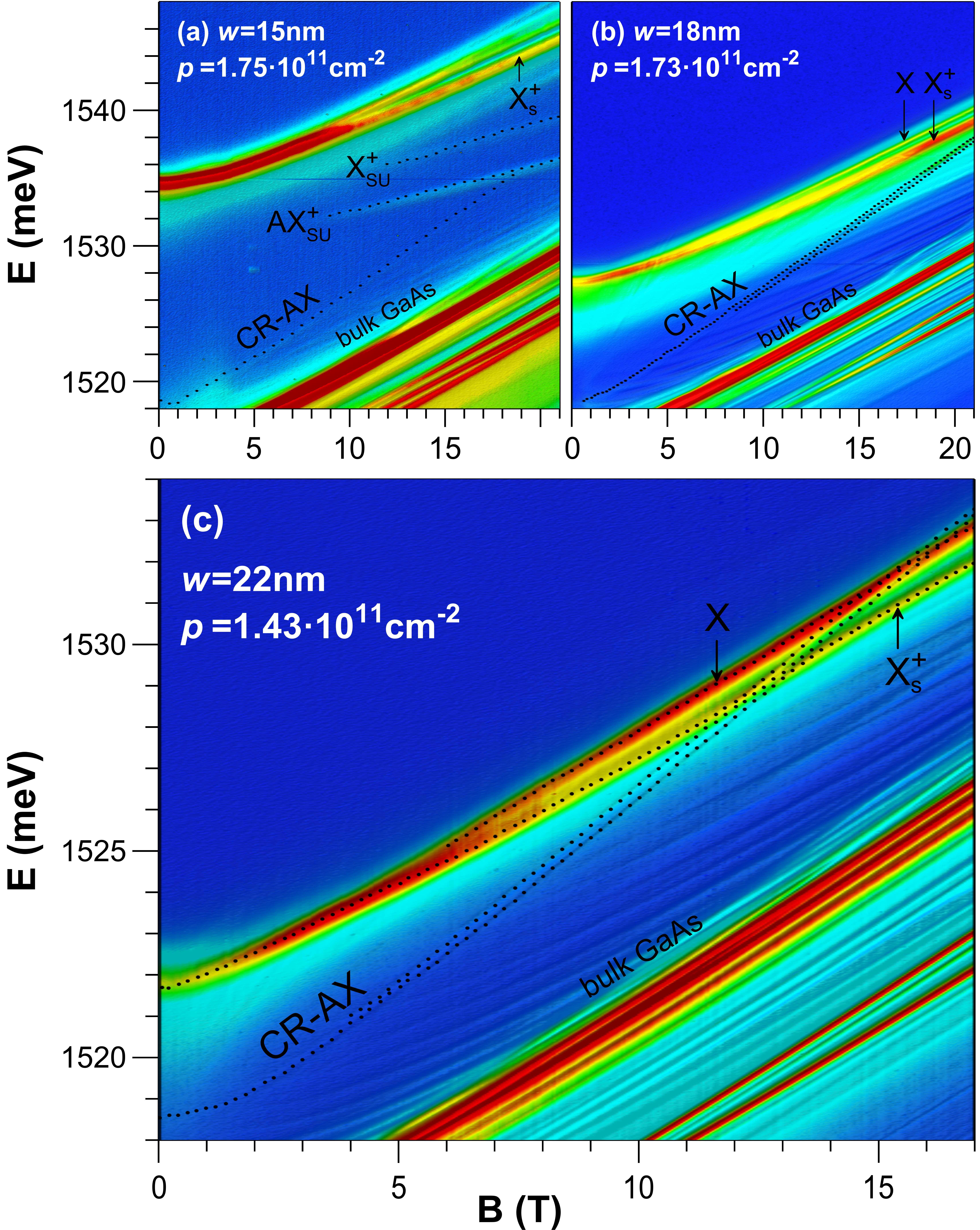}
\caption{(color online) Evolution of the photoluminescence spectra of the narrow symmetric quantum well of width $w=15$~nm (a) and two wider asymmetric wells of widths $w=18$~nm (b) and 22~nm (c) in magnetic field $B$, collected in polarization $\sigma^-$ (higher-intensity), at low temperature $T=1.8$~K, and under laser power density $P=20$~mW/cm$^2$. The identified lines are: exciton (X), positive trion family (X$^+_{\rm s}$, X$^+_{\rm tb}$, and X$^+_{\rm td}$), acceptor-bound complexes (AX and AX$^+$), shake-up lines (SU), and cyclotron replicas (CR). The color scheme is the same as in Fig.~\ref{fig1}.}
\label{fig2}
\end{figure}

\begin{figure}
\includegraphics[width=0.45\textwidth]{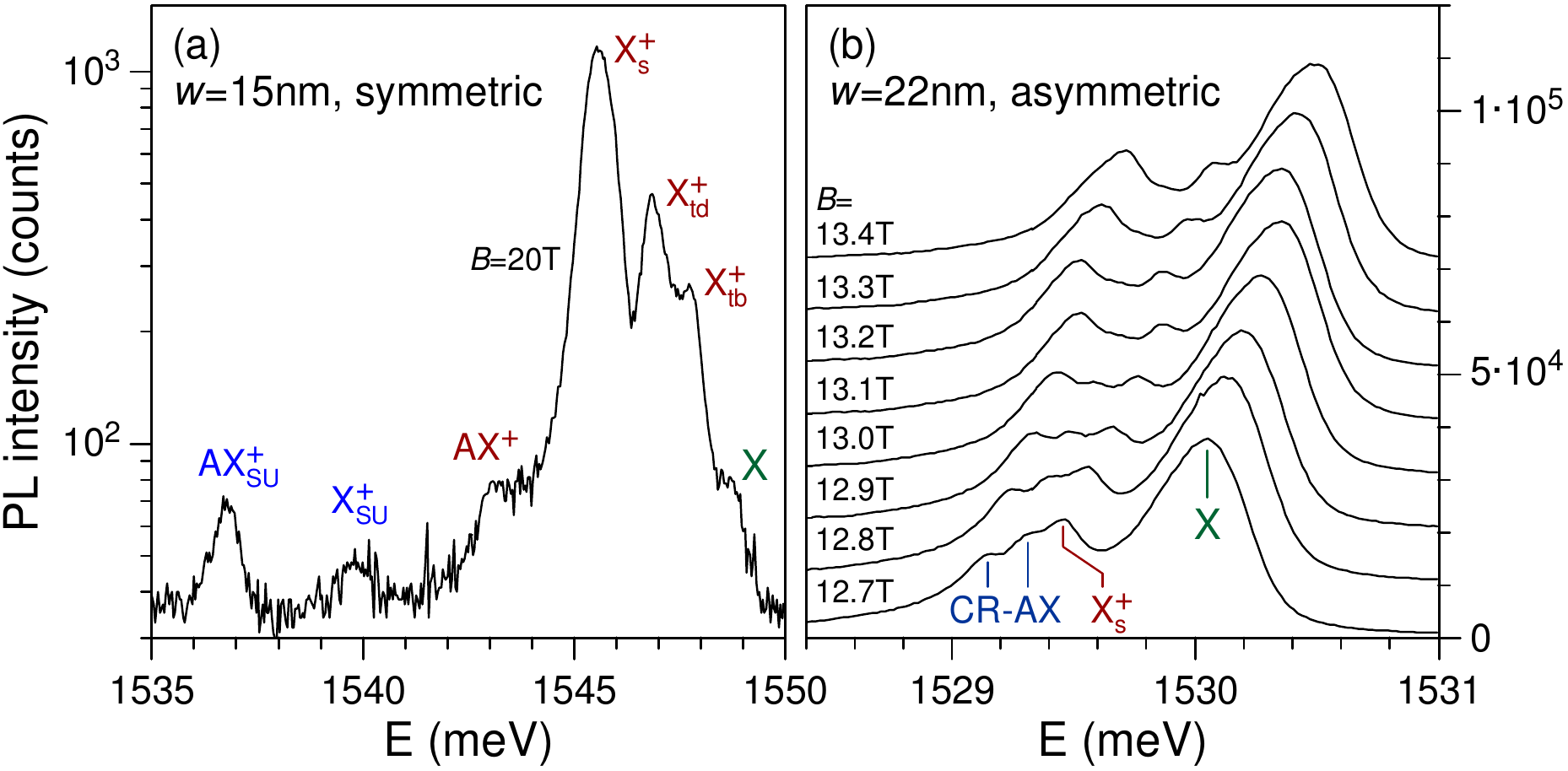}
\caption{(color online) Examples of the $\sigma^-$-polarized high-field photoluminescence spectra of two wells from Fig.~\ref{fig2}: (a) $w=15$~nm, symmetric; (b) $w=22$~nm, asymmetric.}
\label{fig3}
\end{figure}

In Fig.~\ref{fig2} the field evolution of the photoluminescence spectrum in polarization $\sigma^-$ (corresponding to the higher emission intensity) is presented for one symmetric ($w=15$~nm) and two asymmetric wells ($w=18$ and 22~nm). Examples of high-field spectra are shown in Fig.~\ref{fig3}. In each spectrum we have identified in the high-energy sector several strong emission lines, attributed to the nearly free exciton (X) and spin-singlet trion (X$^+_{\rm s}$). The complete trion family \cite{Wojs00,Wojs07}: singlet (X$^+_{\rm s}$), bright triplet (X$^+_{\rm tb}$), and dark triplet (X$^+_{\rm td}$), has been resolved exclusively in the symmetric $w=15$~nm well. 

We use the term ``nearly free'' to signify that although efficient emission from an excitonic complex generally requires some lateral localization \cite{Solovyev09} (by remote charged impurities or well width fluctuations), for a ``nearly free'' state its effect on recombination energy is not substantial. Also distinctive is the effect of localization on the dynamics in the growth direction: the emission energy is sensitive to the well width for the ``nearly free'' states, in contrast to the ``strongly bound'' ones, clung to one side of a well as a result of attraction to an interface acceptor.

Emergence of multiple emission lines from all of these excitonic complexes generally requires sufficient power density and/or magnetic field. Hence, all spectra in Fig.~\ref{fig2} begin at low fields with only one strong line (X or X$^+_{\rm s}$) at high energy. In a sufficiently high field, it splits into two lines, gradually separating in energy with further increase of the field. Their interpretation was aided by realistic numerics \cite{Wojs07} which included effects of finite well width, anisotropic hole mass, Landau level and quantum well subband mixing, and carrier correlations, but excluded light-heavy hole subband mixing \cite{Wieck84} and disorder \cite{Solovyev09}.

On the other hand, the lower-energy part of each shown spectrum contains also multiple weaker lines even for the lowest magnetic fields and weakest excitations. Most of them shift with the magnetic field in parallel to neutral and charged excitons, with the same slope of energy-field dependence, saturating at 0.7~meV/T at higher fields. However, in the spectra collected for higher power densities, we distinguish also two other kinds of lines, with either lower or higher slopes (i.e., either departing from the main lines or approaching them), and interpreted as hole cyclotron replicas of different excitonic complexes.

In particular, in the symmetric $w=15$~nm well we have observed shake-up lines of the nearly free and acceptor-bound positive trions (the latter with the $A^-$ in the well). Their energy-field slope in high fields is 0.4~meV/T. The difference from the main lines matches the hole cyclotron energy determined from the cyclotron resonance of low-density GaAs wells (interpolation of data from Ref.~\cite{Cole97} to $w=22$~nm gives the slope 0.28~meV/T yielding an effective mass $m_{\rm h}=0.38$). Remarkably, these shake-up lines cannot be resolved in the spectra of our wider wells, where they fall inside the emission range of bulk GaAs.

However, it is the other group of lines whose behavior we have found most intriguing. Two of them, denoted as CR-AX$_{\rm s}$ and CR-AX$_{\rm t}$ (as we will justify further) in Fig.~\ref{fig2}, are detected in the spectrum of each well, and remarkably in each one they follow a virtually identical energy-field dependence. In particular, regardless of the well width $w$, at $B=0$ they begin at the same energy of 1518~meV and with the same splitting of 0.3~meV. When magnetic field is switched on, they linearly approach the high intensity lines, with the energy-field slope saturating at 1.0~meV/T in high fields. Comparison of this slope with those of X and X$^+$ yields a field-independent difference of $\gamma=0.3$~meV/T, i.e., exactly opposite to that characterizing the hole shake-up lines.

At this point let us express a few obvious conclusions: (i) The fact that the CR-AX lines repeat unchanged in wells of different width ($w=15$ to 40~nm) precludes their origin in the interior of the well. (ii) The low energy at $B=0$ precludes the origin inside the barrier. Furthermore, the energy position several meV below the free X and X$^+$ states inside the (even fairly wide) well suggests strong localization, most likely on an acceptor. (iii) The energy-field slope (compared to X and X$^+$) reveals recombination accompanied by a hole cyclotron relaxation. The CR-AX transitions are hence related to the previously observed ``combined exciton-cyclotron resonance'' \cite{Yakovlev97}, which was an {\em electron} cyclotron replica of the {\em nearly free} exciton.

The most intriguing effect is observed when the pair of CR-AX lines approach the trion line: both X$^+$/CR-AX$_{\rm s}$ and X$^+$/CR-AX$_{\rm t}$ crossings are clearly avoided. We have only been able to observe this effect in the $w=22$~nm well, where it occurs at an accessible magnetic field $B_c=13.2$~T and the lines are sufficiently sharp (0.4~meV full width at the half-maximum, compared to 0.2~meV thermal broadening). In a narrower $w=18$~nm well, extrapolation of the available data points predicts the prohibitively high anti-crossing field $B_c\approx28$~T. In a wider well of $w=25$~nm, the resonance occurs at $B_c<4$~T, and the anti-crossing is not resolved due to a larger line width in this field range.

The relevant sector of Fig.~\ref{fig2}(c) has been magnified in Fig.~\ref{fig1}, which clearly displays both avoided crossings -- a convincing signature of the mixing of the corresponding states: CR-AX and X$^+$ \cite{initial}. Fig.~\ref{fig1} also makes it clear that the intensities of the CR-AX lines become greatly enhanced in the anti-crossing region -- as a simple consequence of the mixing with the more radiative nearly free trion state. In higher magnetic fields, above the anti-crossing region, the CR-AX and X$^+$ lines resume their initial energy-field dependences and intensities.

As anti-crossing of the CR-AX lines with the X$^+$ line implies coupling between the corresponding radiative states (initial states for recombination), it also indicates spatial proximity of the CR-AX recombination to the quantum well (hosting the X$^+$). Combination of this fact with the earlier conclusions (i--iii) allows us to attribute the CR-AX lines to the cyclotron replica of an excitonic complex bound to an acceptor at the interface separating the GaAs well from its GaAlAs barrier. Emission from such a bound AX complex and its shake-up AX$_{\rm SU}$ have been detected at lower fields in an undoped quantum well \cite{Volkov98}, but the CR-AX line (requiring the hole gas) to the best of our knowledge has not been previously observed.

That the relevant bound excitonic state is neutral rather than charged follows from the exact course of emission lines near the anti-crossing. The hypothetical charged complex AX$^+$ would consist of a bare interface acceptor (negative point charge), one electron, and three holes. Previous calculations \cite{Bryja07} hint that its ground state would have a doublet spin configuration, corresponding to the unpolarized holes (total three-hole spin of 1/2; we adopt a convention assigning spin projections $\pm1/2$ to the ``up'' and ``down'' heavy hole states), with the quadruplet configuration (polarized holes) excited by several meV. Thus, only the doublet state AX$^+_{\rm d}$ would be populated at low temperatures. In the relevant polarization $\sigma^-$ it would have two recombination channels, ending in a singlet or triplet state of the charged acceptor (AX$^+_{\rm d}$ $\rightarrow$ A$^+_{\rm s}$ or A$^+_{\rm t}$, where A$^+$ has two holes bound to a bare acceptor). Thus, its emission would be split by a small exchange energy gap in the final state (A$^+$; in the optically active angular momentum channel $M=-1$ \cite{Bryja07}). At first sight, this agrees with the observed 0.3~meV splitting of the CR-AX lines, but the exact course of the anti-crossing lines turns out incorrect.

The following alternative scenario involves a neutral complex AX, which has an electron and two holes (all inside the well) bound to a bare interface acceptor A$^-$. Its two spin configurations of the pair of holes, singlet (AX$_{\rm s}$) and triplet (AX$_{\rm t}$), are separated by an exchange gap of a fraction of meV. Both configurations should be populated in experiment, and each would have a single $\sigma^-$-polarized recombination channel, ending in the neutral acceptor A (one hole bound to a bare acceptor). Thus, emission from a neutral complex is split by a singlet-triplet exchange in the initial state, naturally explaining the observed 0.3~meV splitting of the CR-AX lines. As we discuss below, it also correctly predicts details of the anti-crossing pattern.

\begin{figure}
\includegraphics[width=0.45\textwidth]{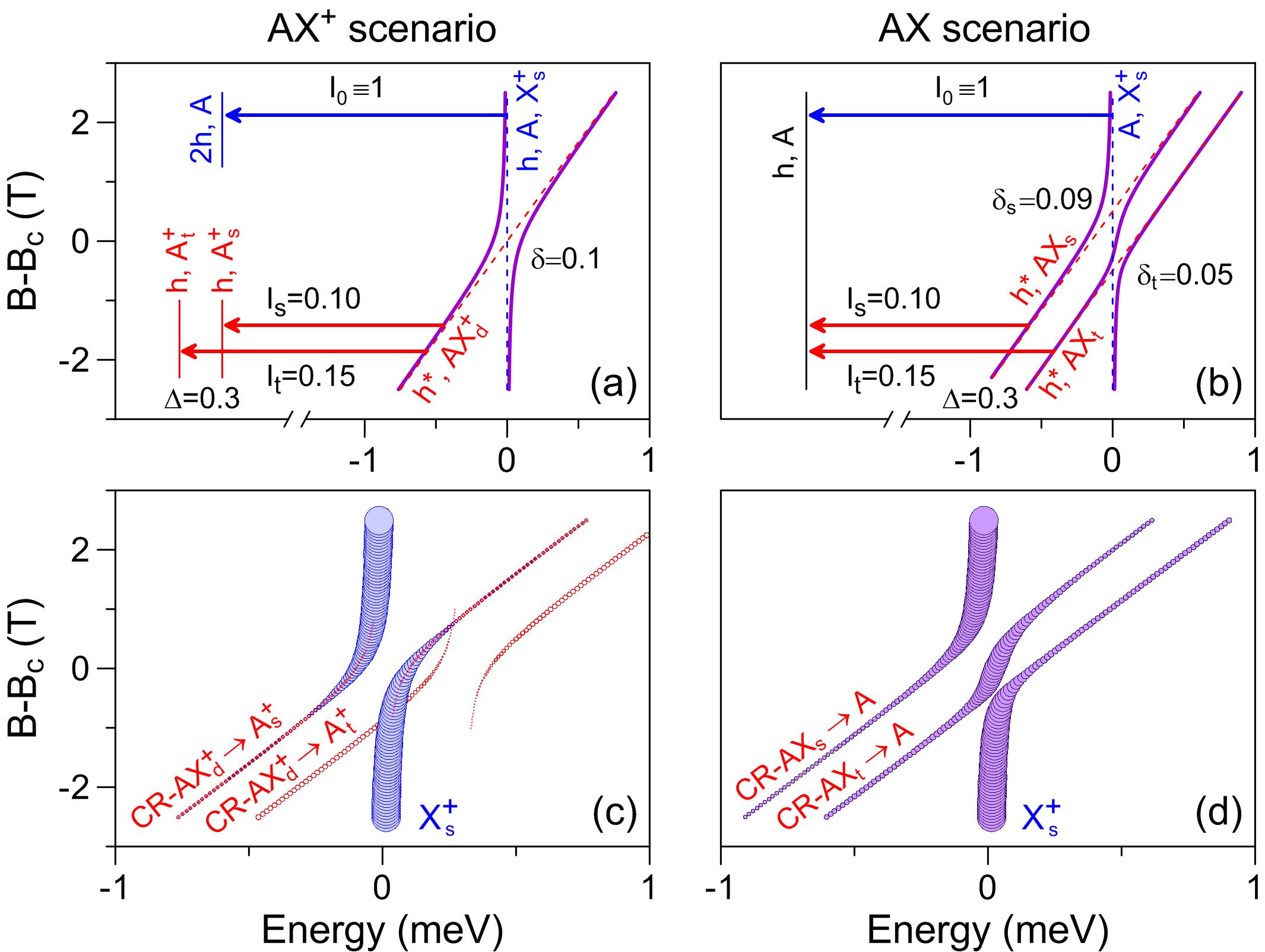}
\caption{(color online) Top: Energy-field dependence for the initial and final states of two candidate scenarios for the observed anti-crossing pattern of Fig.~\ref{fig1} (left/right -- cyclotron replica of a charged/neutral acceptor-bound excitonic complex mixed with a free trion, with the two-hole exchange splitting in the final/initial configuration). Magnetic field $B$ is counted from the resonance field $B_c$, and the energy -- from the trion position at resonance. Exchange and coupling energies ($\Delta$ and $\delta$) are quoted in meV. Oscillator strengths away from resonance ($I$) are defined relative to the free trion. Bottom: Corresponding field-dependences of the emission energy and intensity (indicated by dot radii). Comparison with Fig.~\ref{fig1} allows positive identification of the CR-AX transition.}
\label{fig4}
\end{figure}

Let us now compare in more detail these two scenarios, involving either a charged or a neutral excitonic complex. In each case, the recombination event is accompanied by a cyclotron relaxation of a nearby hole, thermally excited to a higher Landau level. In a particular field, when the hole cyclotron energy equals the binding energy of the AX complex, the cyclotron transition brings to resonance the free and localized states, thus allowing their efficient coupling (by an exciton transfer). Fig.~\ref{fig4} presents the energy-field dependence for the relevant initial and final states and the connecting transitions, obtained in a simple phenomenological model \cite{model}. The exchange gap $\Delta$ and the relative intensities $I$ of the unmixed transitions were read from the experimental spectra (away from the crossings). For the ``charged'' scenario (left), degeneracy of the two final states $h+{\rm A}^+_{\rm s}$ and $2h+{\rm A}$ (i.e., zero ${\rm A}^+_{\rm s}$ binding energy in the relevant $M=-1$ channel), is needed (and, actually, justified by the previous calculation \cite{Bryja07}) to maximally reduce the predicted anti-crossing pattern. Yet, regardless of the coupling constant $\delta$, the experimental behavior cannot be reproduced, as it is shown in Fig.~\ref{fig4}(c) for $\delta=0.1$~meV. For the ``neutral'' scenario (right), a single, common final state $h+{\rm A}$ removes the need of additional degeneracy. Moreover, assuming two coupling constants $\delta=0.05$ and 0.09~meV (distinguished by spin configuration of the AX) produces a convincing match with experiment, which is the basis for our interpretation (and notation) of the CR-AX lines.

\begin{figure}
\includegraphics[width=0.45\textwidth]{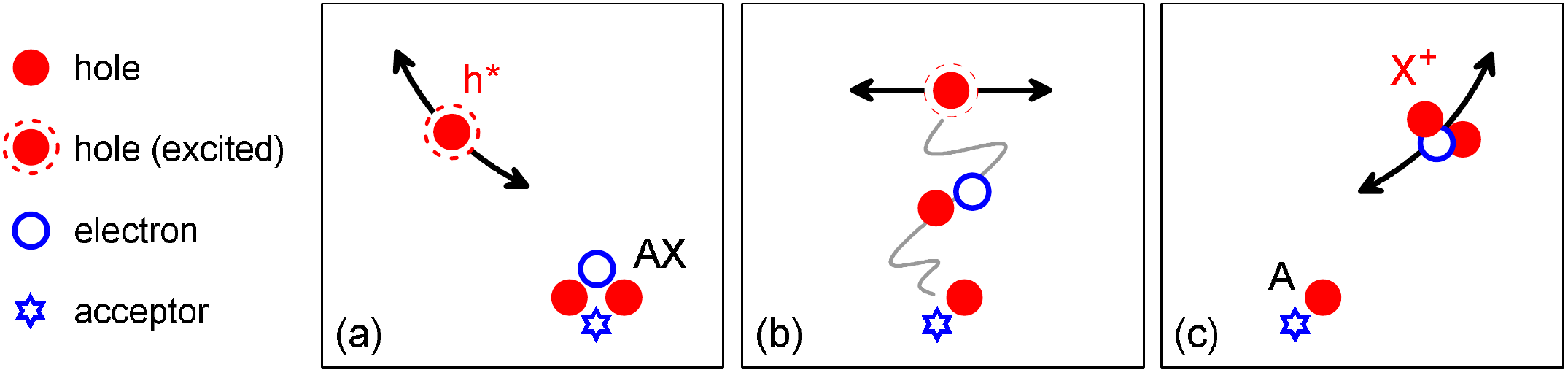}
\caption{(color online) Schematic picture of the coupling between the acceptor-bound exciton AX and the nearly free trion X$^+$ through the exchange of an electron-hole pair combined with a hole cyclotron transition.}
\label{fig5}
\end{figure}

Having determined the relevant radiative states $h^*+{\rm AX}$ and ${\rm A}+{\rm X}^+$, let us now discuss their coupling mechanism. The responsible process is the transfer of an electron-hole pair, as illustrated in Fig.~\ref{fig5}: A hole thermally excited to a higher Landau level ($h^*$) approaches an AX (in the singlet or triplet spin state). When their constituents engage in the Coulomb interaction, the hole relaxes to the lowest Landau level, releasing a quantum of cyclotron energy ($\hbar\omega_c\approx4$~meV at the resonant magnetic field $B_c=13.2$~T) which is used to unbind an $e$-$h$ pair from the acceptor (breaking up the AX) and bind it to the hole (forming an X$^+$). Of course, the reverse process is equally possible, beginning with an X$^+$ approaching an A, and ending with the $h^*$ leaving the AX. Remarkably, different localization of the transferred $e$-$h$ pair in the mixed configurations will make the coupling strength $\delta$ sensitive to the actual extent of the nearly free state and/or to the average distance between the acceptors.

Finally, Fig.~\ref{fig2}(c) also reveals anti-crossing of the CR-AX lines with the free exciton. This is straightforward to explain in light of the above, the only difference being that a larger cyclotron energy $\hbar\omega_c\approx4.8$~meV at the higher field $B_c\approx16$~T brings the AX in resonance with the exciton instead of a trion: $h^*+{\rm AX}\leftrightarrow h+{\rm X}+{\rm A}$.

In conclusion, in magneto-photoluminescence spectra of positively doped quantum wells we have demonstrated coupling between the nearly free and strongly localized excitonic complexes, X/X$^+$ and AX, brought in resonance at a particular high magnetic field by an additional hole cyclotron transition. Coupling occurs through the exchange of an electron-hole pair. The result is a complex magnetically tunable coherent state whose optical spectrum probes weak localization in the quantum well.

A.W. and L.B. thank W. Bardyszewski for discussions. Work supported by Polish MNiSW grant N202179538 (L.B. and J.J.), EU Marie Curie grant PCIG09-GA-2011-294186 (A.W.), BMBF-QuaHL-Rep and DFG grants (G.B., D.R.Y., M.B., D.R., and A.D.W.), and EuroMagNET II under EU Contract No. 228043 (P.P. and M.P.).


\begin{thebibliography}{99}
\bibitem{initial}
An alternative field-induced crossing of the final states (nearly free and strongly localized hole) does not occur.
\bibitem{model} 
The initial configurations in Fig.~\ref{fig4} are described by model 2- and 3-level Hamiltonians determined by the CR-AX splitting $\Delta=0.3$~meV, the relative CR-AX/X$^+$ energy-field slope $\gamma=0.3$~meV/T, and the indicated adjustable coupling constants $\delta$.
\end{thebibliography}
\end{document}